# Development of five multifibre links for the OPTIMOS-EVE study for the E-ELT

Isabelle Guinouard*[a], Fanny Chemla[a], Hector Flores[a], Jean-Michel Huet[a], François Hammer[a], Gerben Wulterkens[b]

[a] GEPI, Observatoire de Paris, 5 Place Jules Janssen, F-92195 Meudon Cedex, France;
[b] Radboud University Nijmegen, P.O. Box 9010, Nijmegen, Netherlands

## ABSTRACT

The OPTIMOS-EVE concept provides optical to near-infrared (370-1700 nm) spectroscopy, with three spectral resolution (5000, 15000 and 30000), with high simultaneous multiplex (at least 200). The optical fibre links are distributed in five kinds of bundles: several hundreds of mono-object systems with three types of bundles, fibre size being used to adapt slit with, and thus spectral resolution, 30 deployable medium IFUs (about 2"x3") and one large IFU (about 6"x12").

This paper gives an overview of the design of each mode and describes the specific developments required to optimise the performances of the fibre system.

## 1 INTRODUCTION

The instrument consists of three main sub-systems: a pick-and-place positioner, fibre bundles for various spectral resolutions and integral field units and two highly efficient VIS-NIR spectrographs with VPH gratins working in 1$^{st}$ order.

The fibre system was designed by the GEPI at Observatoire de Paris.

## 2 DESCRIPTION

A schematic view of this system is shown in Figure 1. It is designed to fit within the volume and mass limits of the focal plane station.

* Contact: Isabelle.Guinouard@obspm.fr; phone +33-1-45077983; fax +33-1-45077709. GEPI, Observatoire de Paris

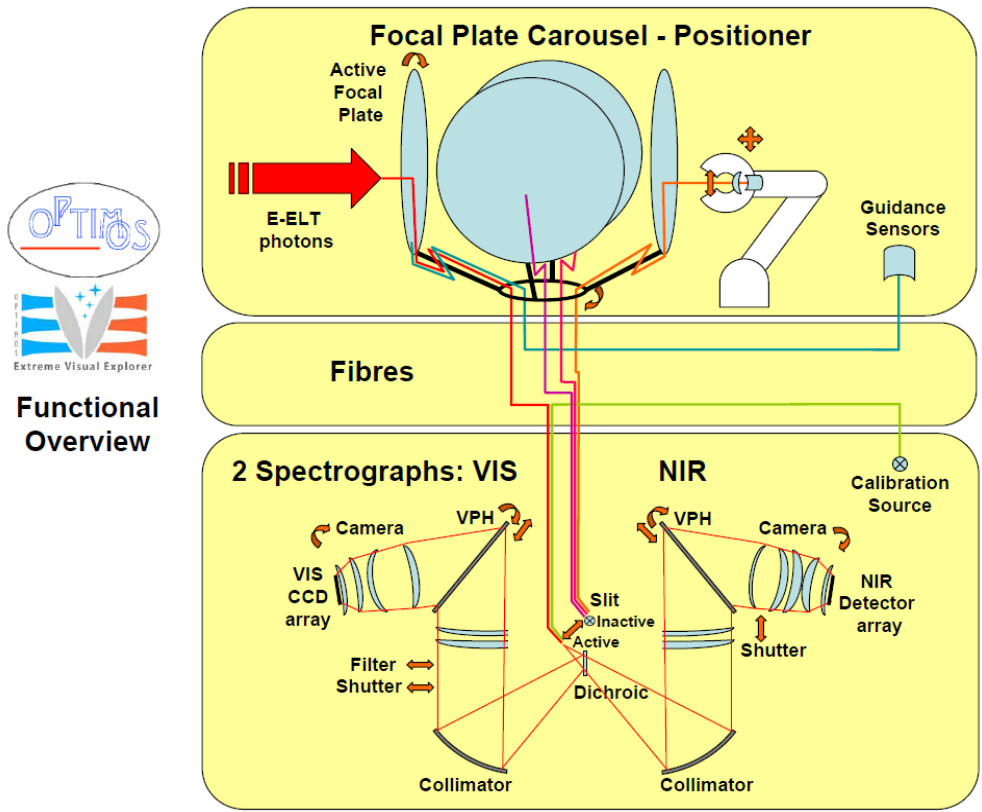

Figure 1 Schematic overview of OPTIMOS-EVE

The OPTIMOS-EVE fibre system ensures the link between the spectrographs and the positioner. This fibre system will be composed of five kinds of links: 3 types of Mono-Object (MO), 1 type of Medium-IFU (MI) and 1 Large-IFU (LI). For these five modes, the optical aperture conversions at the input end on the bundle are realized by a coupling with a microlens. The pupil of the telescope is imaged on the fibre core.

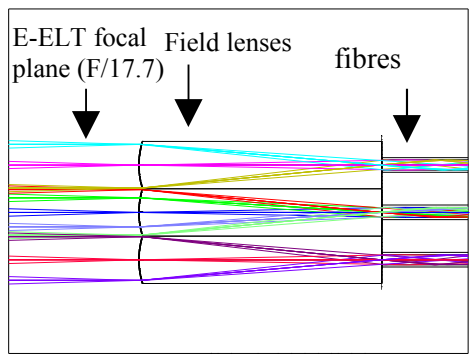

Figure 2 Fibre injection principle (pupil image on fibre core)

The MO field of view is 0.9'' for Low and Medium Resolution (LR and MR) and 0.81'' in High Resolution (HR). The Integral fiel Unit MI and LI have 0.3'' spatial sampling on sky and operate in Low Resolution (LR) only.

At the output of the bundle the fibres of one type of fibre buttons, are arranged into two pseudo slits, feeding into the spectrographs. For the output no microlenses are required since the spectrograph collects an aperture of f/3.5.

The table below summarizes the sampled strategies adopted by OPTIMOS-EVE.

Table 1. Multiplex, spectral resolution, aperture and on sky sampling for the 5 observing modes.

| Observing Mode | Multiplex | Spectral resolution | Aperture | Microlens sampling on sky |
|---|---|---|---|---|
| MO-LO | 240 | 5000 | 0.9'' | 0.3'' |
| MO-MR | 70 | 15000 | 0.9'' | 0.18'' |
| MO-HR | 40 | 30000 | 0.81'' | 0.09'' |
| MI-LR | 30 | 5000 | 1.8''x3'' | 0.3'' |
| LI-LR | 1 | 5000 | 7.8''x13.5'' | 0.3'' |

# 3 MONO-OBJECT LOW RESOLUTION (MO-LR)

This mode presents a high multiplex capacity with a maximum of 240 MO-LR buttons. Each button consists of a bundle of 7 fibres (each 0.3'' diameter) and combined they cover a 0.9'' circular input aperture on the sky as illustrated below. The size of a single fibre core is 223μm.

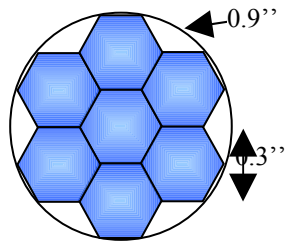

*Figure 3 MO-LR design*

The MO-LR mode has to be duplicated in order to have two focal plates, one observing while the other is being reconfigured.

**3.1 MO-LR conceptual design**

The MO-LR fibres will be arranged at output into a pseudo-slit that is part of the slit positioner mechanism in the spectrograph. One pseudo slit will be able to accommodate 120 MO-LR buttons, 110 of which are made up of the 7 fibres from one MO-LR button and 10 are made up of the 7 fibres of one MO-LR button plus an additional simultaneous wavelength calibration fibre. Two slit units will feed into the two spectrographs to cover all 240 MO-LR buttons at the same time.

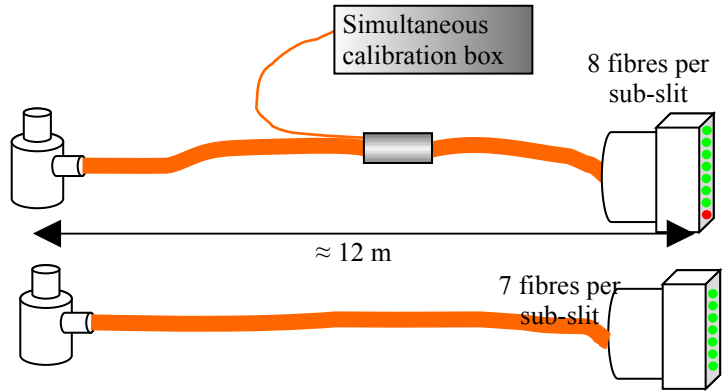

Figure 4 MO-LR conceptual schematic drawing

# 4 MONO-OBJECT MEDIUM RESOLUTION (MO-MR)

This mode presents a maximum of 70 MO-MR buttons. Each button consists of 19 fibres (each 0.18" diameter) that, combined, cover a circular input aperture on the sky (of 0.9") similar to the MO-LR but now providing higher spectral resolution. The size of each fibre core is 134 μm. The MO-MR mode has to be duplicated in order to have two focal plates, one observing while the other is being reconfigured.

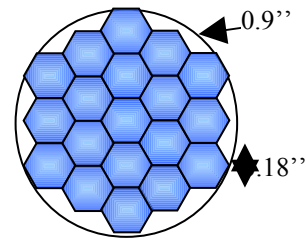

Figure 5 MO-MR design

## 4.1 MO-MR conceptual design

The arrangement of the 70 MO-MR buttons at the input stage to the spectrographs is very similar to that of the MO-LR. Each button is combined in a subslit and adding up to 35 MO-MR buttons per slit, 10 subslits will have a calibration fibre attached for simultaneous calibration. Two slits (each carrying 35 MO-MR buttons) feed the two separate spectrographs.

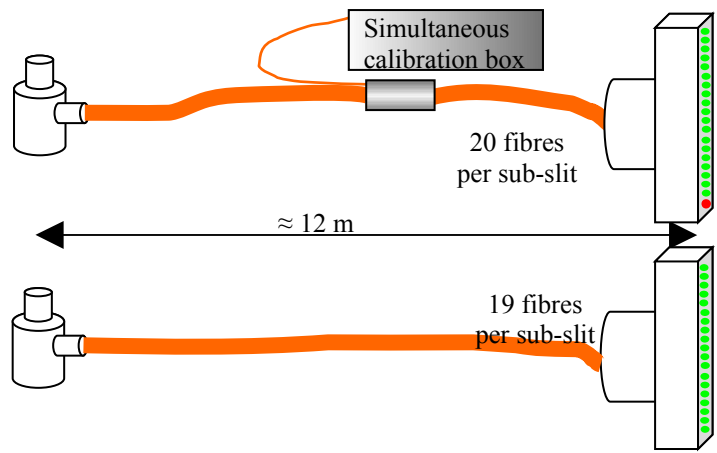

Figure 6 MO-MR conceptual schematic drawing

# 5 MONO-OBJECT HIGH RESOLUTION (MO-HR)

The design of the MO-HR buttons is conceptually different from that of the MO-LR and MO-MR sets. In the MO-HR button the circular input aperture on the sky of 0.81" is fed into a 7-fold array of 0.27" microlenses. At the output of each subsequent fibre (core diameter 200 μm) a 7 microlens array of each 0.09" dividing the aperture into 7 subapertures. So the 0.81" input aperture is thus divided in 7x7= 49 sub-apertures of 0.09", to achieve the high resolution required. The diameter of these final fibres is 67 μm.

The reason why we have not chosen to put the 0.09'' microlenses directly in the focal plane like in other modes is that the focal length is shorter than the aperture diameter, making it impossible to fold the button with a 90° prism. Also, having a relay in the middle of fibre allows to increase scrambling in the fibre, and thus to improve radial velocity accuracy with a more homogeneous flux at 0.09'' fibre output.

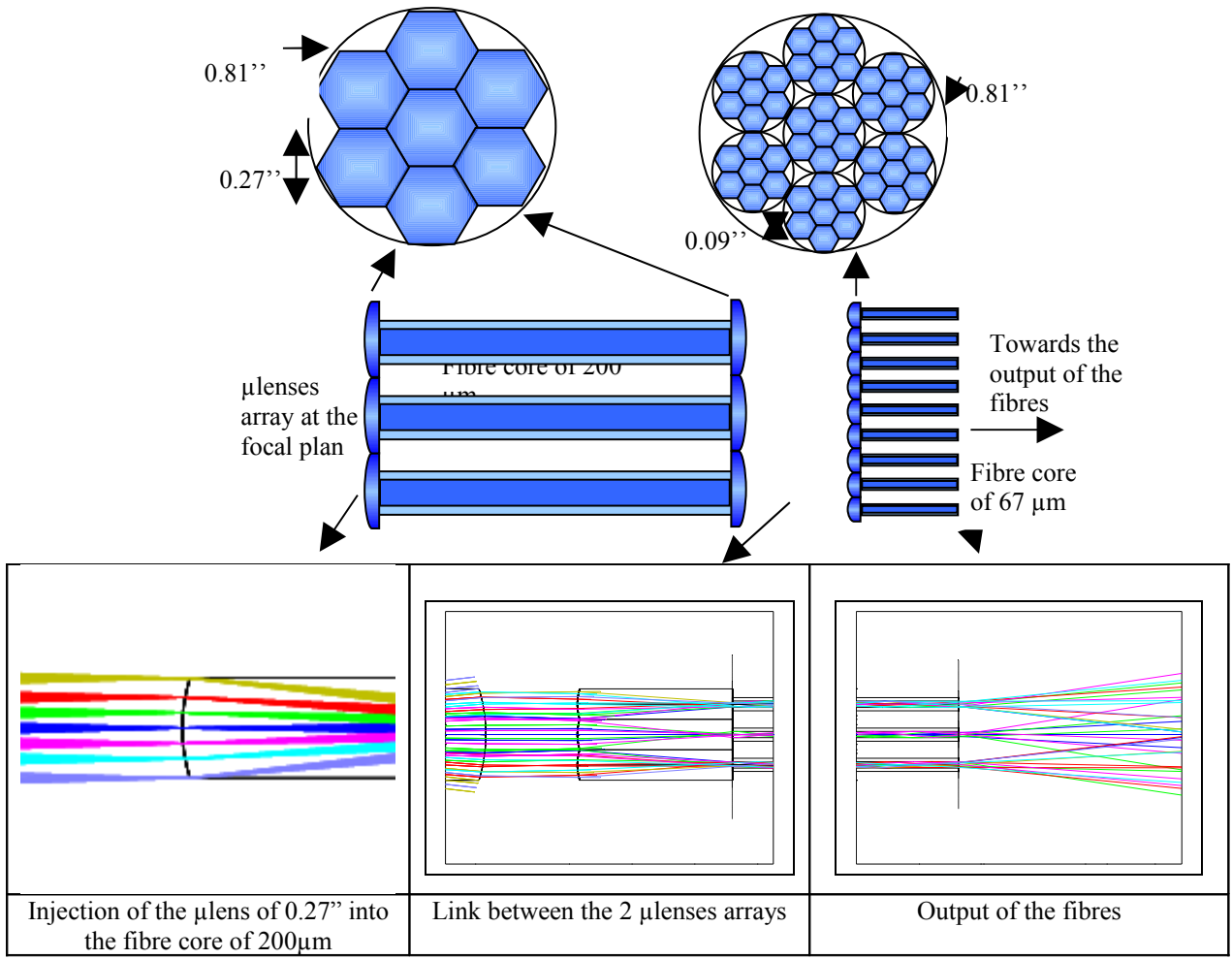

*Figure 7 MO-HR design*

### 5.1 MO-HR conceptual design

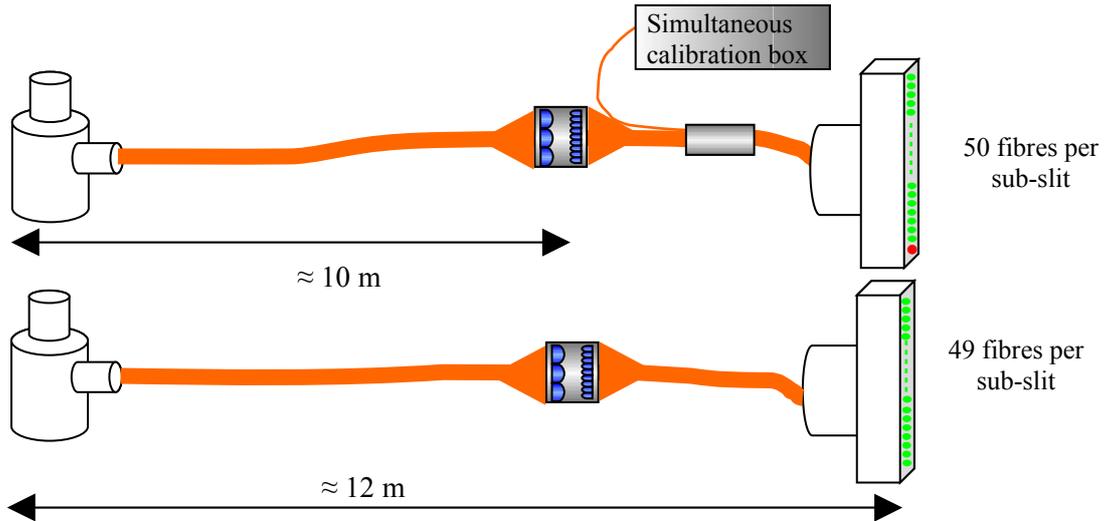

Figure 8 MO-HR conceptual schematic drawing

## 6 MEDIUM-IFU (MI-LR)

Thirty Medium IFUs can be positioned anywhere over the 10 arcminute field of view. The 30 Medium-IFUs (MI-LR) consist each of 52 target fibres (each 0.3" in diameter) and 4 outlying sky fibres, at a fixed position with respect to the science fibres, in an arrangement shown below. The total sky coverage per MI-LR is 1.8"x3".

The arrangement of the fibre bundles into the spectrographs is again similar as for the MO buttons. A pseudo slit is made out of the fibres in subslits, 10 of which are equipped with an additional calibration fibre.

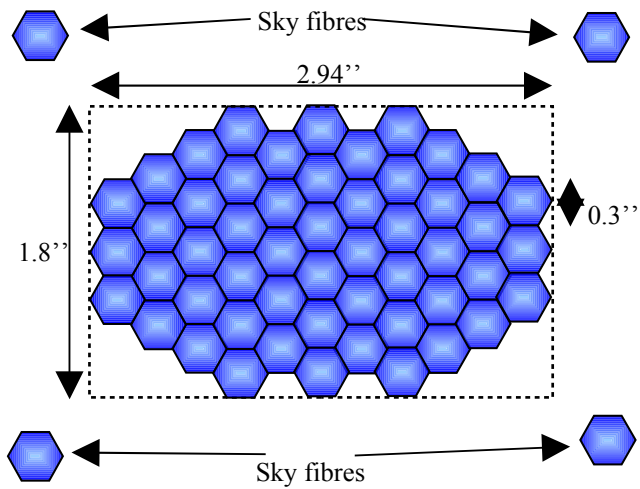

Figure 9 Medium IFU design

## 7 LARGE-IFU (LI-LR)

There is one large IFU. The architecture is similar to that of the medium-IFU, but with a 13.5'' x 7.8'' aperture on the sky made possible by 1560 fibres, each 0.3" diameter. There are 90 sky fibres located at fixed positions around the LI-LR.
Only one focal plate of the positioner will be equipped with the LI-LR. The location of the LI-LR is fixed.
The 1560 object fibres and the 90 sky fibres will be distributed over the 2 combined visible and NIR spectrographs. So there are 780 object fibres and 45 sky fibres on each slit.

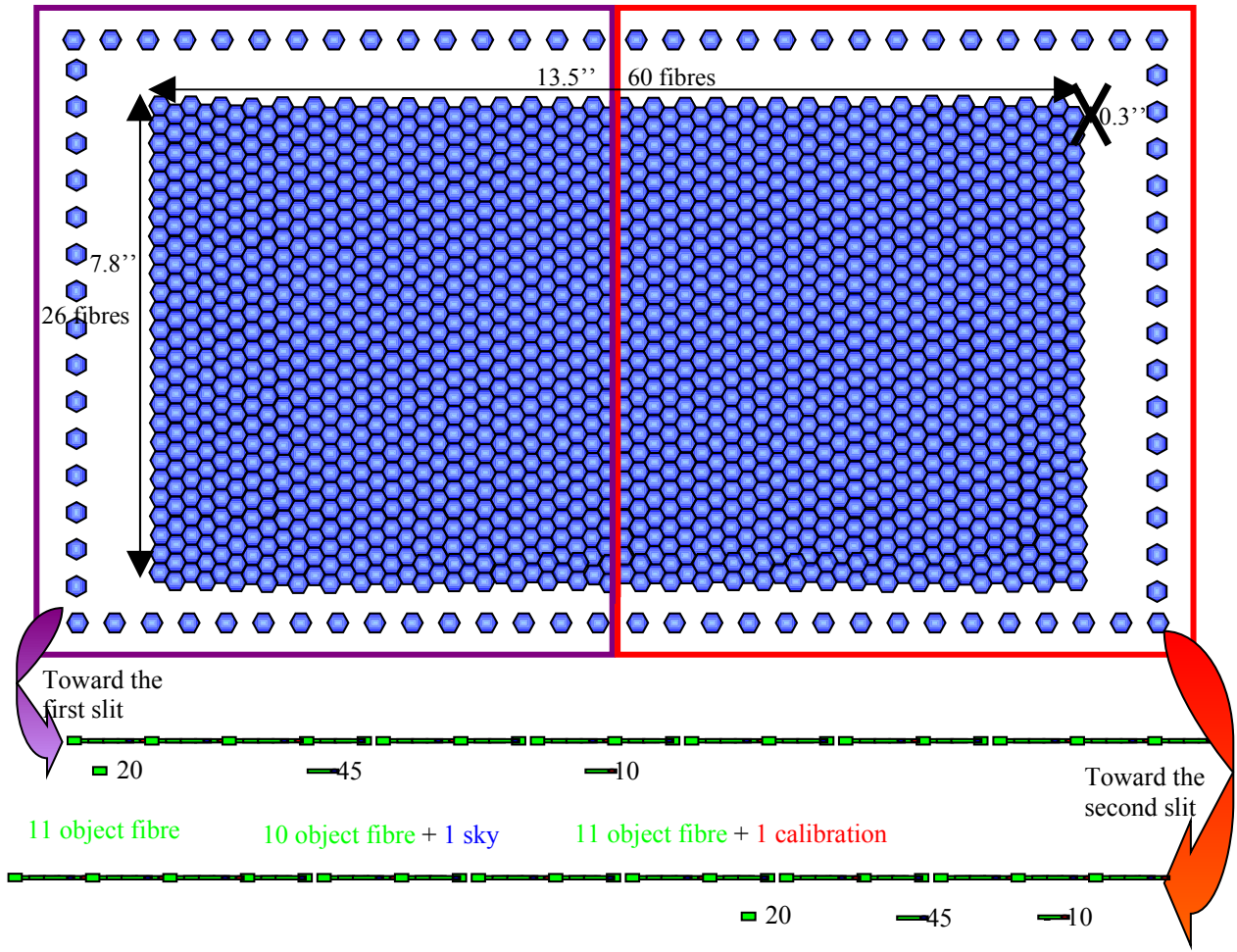

*Figure 10 LI-LR conceptual schematic drawing*

## 8 FIBRES DESCRIPTION

**8.1 Fibre length**

The length of the fibres will be determined by the distance between the focal plates and the spectrographs. In the current design the spectrographs are directly below the positioner and the distance is minimized. However, additional length is needed for strain relief and rotation of both the focal plate positioner and the slit carriage in the spectrograph. A fibre length of 12 m is sufficient.

**8.2 Fibre transmission**

In the Phase A study only Polymicro fibres are investigated. Polymicro fibres are well known and are used in various astronomical instruments. The transmission curves of four types of Polymicro fibres are shown below. Since all fibres must cover the full wavelength range the choice is limited.

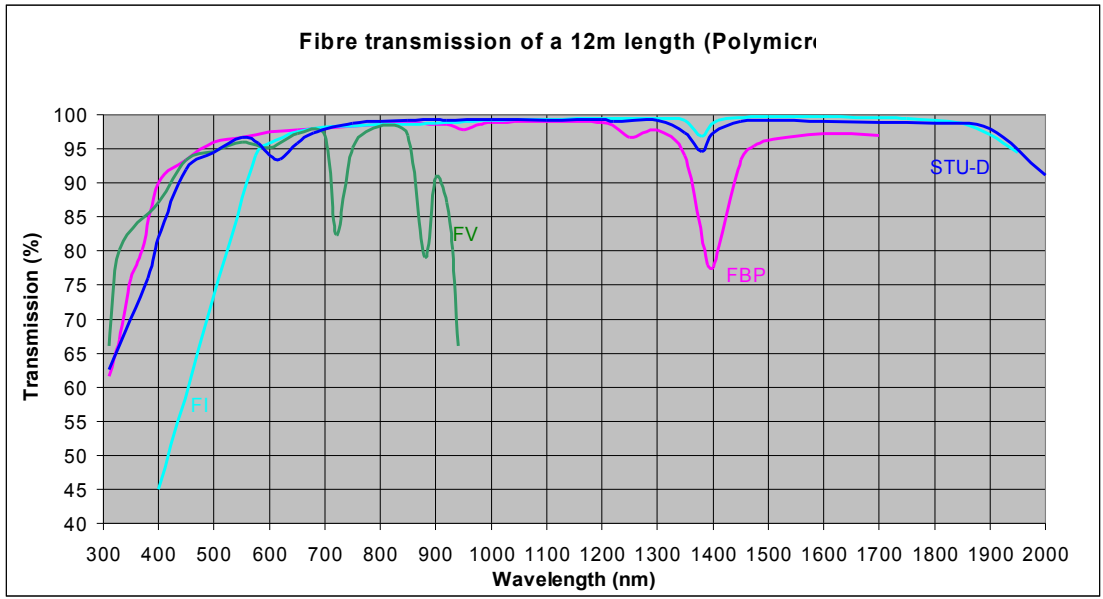

*Figure 11 Fibre intrinsic transmission curves*

Based on the improved UV/Blue transmission as well as the cost the FBP fibres are chosen (FBP is only half the price of STU-D). The extra absorption of FBP at 1400 nm (due to water impurities) is not problematic since this falls right on top of the atmospheric transmission gap between the J and the H-band (also due to water absorption).

**8.3 Fibre structure**

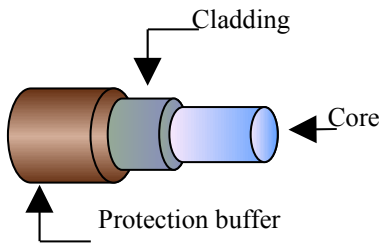

∅ MO LR : 223/267/300µm
∅ MO MR : 134/160/180µm
∅ MO HR : 67/80/95µm
          200/220/240µm
∅ MI/LI : 223/267/300µm

∅ core / cladding / prot.buffer

*Figure 12 Fibre structure*

**8.4 Focal Ratio degradation**

Focal Ratio Degradation (FRD) is the decrease in focal ratio (decrease in effective F-number) in an optical fibre. The ability of a fibre to preserve the angular distribution of the input beam from the telescope to the spectrograph is very important.

The major causes of FRD are mechanical variations in the fibre dimensions with length (under the manufacturer's control) and the mechanical set-up of the instrumentation (under the control of the user). Small variations in the fibre core diameter or core-clad interface can cause mode stripping, resulting in FRD. Both macrobending and microbending will cause FRD.

In OPTIMOS-EVE we inject at a fast F ratio of F/3.65, which permits to limit the FRD. At the output, the F ratio is slightly degraded and is at F/3.5.

# 9 MECHANICAL DESCRIPTION

## 9.1 Entrance

The mechanical quality of the end of the fibres is a well known determining factor in the photometric quality of the fibre. Special attention will therefore be given to the treatment of the end surfaces, and e.g. their link to the microlenses and the parallelism of the output fibres.
The fibres will be holded in a holes array. The bundle of fibres and the folded microlens will be integrated in a magnetic button which be placed in the focal plate.

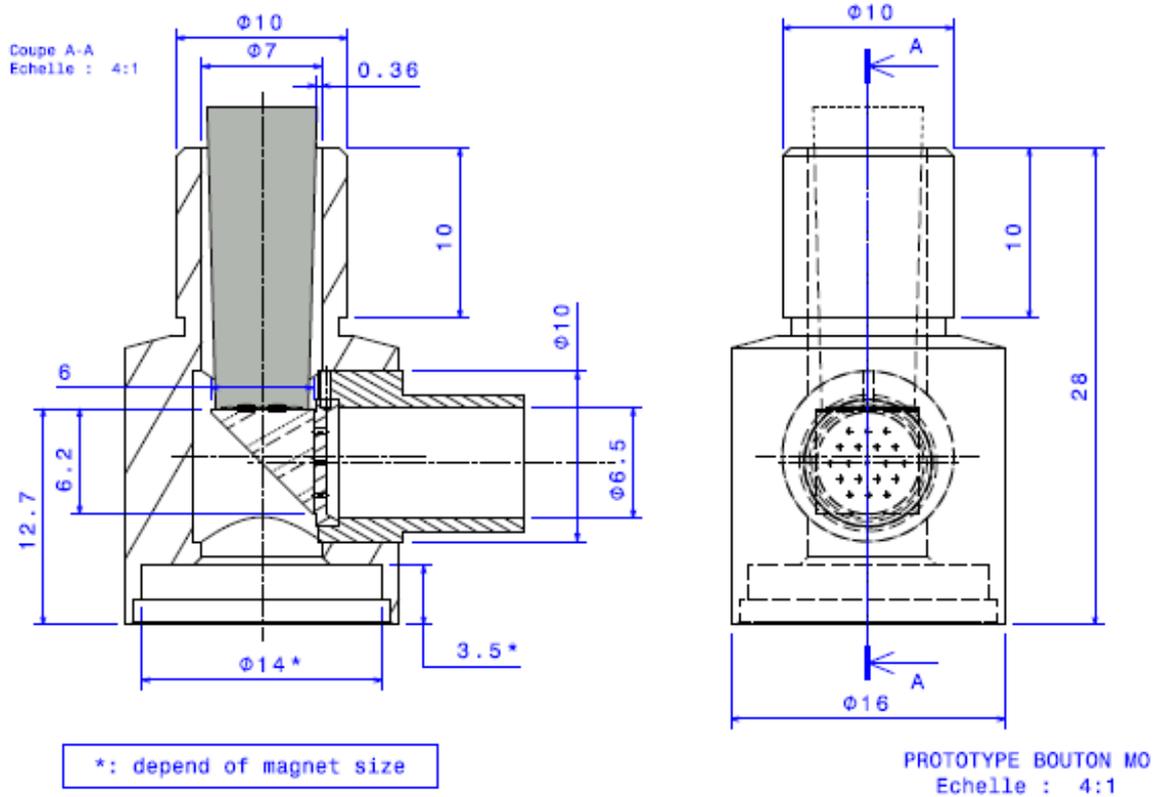

*Figure 13 Design of MO button*

## 9.2 Output

The total height of the slit is 338.7mm. The slit support has to be designed in order to fit the optical constraints of the spectrographs. The fibres inside the spectrograph enclosure are cooled at 193K.

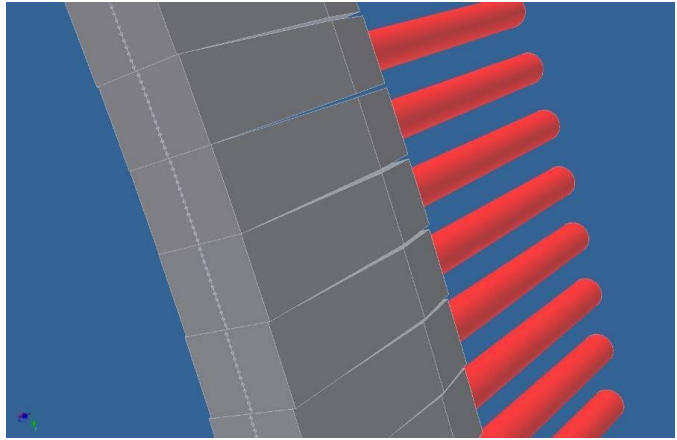

*Figure 14 Entrance slit area*

## 10 CONCLUSION

The optical fibre system for the OPTIMOS-EVE study has been designed by GEPI at Observatoire de Paris. We relied on the experience gained on GIRAFFE for FLAMES.